\documentclass[aps,prd,12pt,amssymb]{revtex4}

\begin{document}

\baselineskip=0.60cm

\newcommand{\ini}{\begin{equation}}
\newcommand{\fin}{\end{equation}}
\newcommand{\inir}{\begin{eqnarray}}
\newcommand{\finr}{\end{eqnarray}}

\def\ol{\overline}
\def\pa{\partial}
\def\ra{\rightarrow}
\def\ts{\times}
\def\df{\dotfill}
\def\bs{\backslash}
\def\dg{\dagger}
\def\la{\lambda}

$~$

\vspace{2 cm}

\title{Neutrino mass and baryogenesis}

\author{D. Falcone}

\affiliation{Dipartimento di Scienze Fisiche,
Universit\`a di Napoli, Via Cintia, Napoli, Italy}

\begin{abstract}
\vspace{1cm}
\noindent
A brief overview of the phenomenology related to the seesaw mechanism and
the baryogenesis via leptogenesis is presented.
\end{abstract}

\maketitle

\newpage

A breakthrough in particle physics occurred in year 1998, when the SuperKamiokande
Collaboration announced evidence for oscillation of atmospheric neutrinos
$\nu_{\mu}$ \cite{sk}. Then, in 2002, at the Sudbury Neutrino Observatory, evidence
for flavor conversion of solar neutrinos $\nu_{e}$ has been found \cite{sno},
pointing to oscillation of solar neutrinos too. These two important results come
after a long series of experiments, back to the atmospheric neutrino anomaly
and the solar neutrino problem. The full terrestrial experiments K2K and KamLAND
confirm the above results for $\nu_{\mu}$ and  $\nu_{e}$, respectively.

$~$

The most natural explanation of neutrino oscillations is that neutrinos have
masses, and leptons mix just (almost) like quarks do. In this case, neutrino
mass eigenstates $\nu_i$ are related to neutrino flavor eigenstates
$\nu_{\alpha}$ by the unitary transformation
$\nu_{\alpha}=U_{\alpha i}\nu_i$, where U is the lepton mixing matrix
\cite{mns}.

$~$

Oscillation experiments provide square mass differences and mixing angles.
One has
\ini
m_3^2-m_2^2=\Delta m^2_{atm} \simeq (0.05 ~\text{eV})^2
\fin
\ini
m_2^2-m_1^2=\Delta m^2_{sol} \simeq (0.009 ~\text{eV})^2
\fin
and the mixing matrix is given by
\ini
U \simeq
\left( \begin{array}{ccc}
\frac{2}{\sqrt6} &  \frac{1}{\sqrt3} & \epsilon \\
-\frac{1}{\sqrt6} &  \frac{1}{\sqrt3} &  \frac{1}{\sqrt2} \\
\frac{1}{\sqrt6} &  -\frac{1}{\sqrt3} &  \frac{1}{\sqrt2} 
\end{array} \right),
\fin
with $\epsilon < 0.2$, where
$U_{e 2} \simeq \frac{1}{\sqrt3}$ is related to solar oscillations and 
$U_{\mu 3} \simeq \frac{1}{\sqrt2}$ to atmospheric oscillations.

$~$

Moreover, from cosmology (Large Scale Structure and Cosmic Microwave Background)
we get
\ini
\sum m_i \lesssim 0.6 ~\text{eV},
\fin
from the tritium beta decay
\ini
U_{e i}^2 m_i^2 \lesssim (2 ~\text{eV})^2,
\fin
and from the neutrinoless double beta decay
\ini
U_{e i}^2 m_i \lesssim 1 ~\text{eV}.
\fin

$~$

Since $m_3^2-m_2^2 \gg m_2^2-m_1^2$, even if the hierarchy is not so strong,
we have three main spectra for the effective neutrinos \cite{alt},
namely the normal
\ini
m_1 < m_2 \ll m_3,
\fin
the inverse
\ini
m_1 \simeq m_2 \gg m_3,
\fin
and the quasi-degenerate
\ini
m_1 \simeq m_2 \simeq m_3.
\fin

$~$

From the summary of experimental informations we note that the neutrino
mass is very small with respect to quarks and charged leptons. Moreover,
about the three independent mixings, $U_{\mu 3}$, $U_{e 2}$, and $U_{e 3}$,
we have $U_{\mu 3}$ maximal, $U_{e 2}$ large but not maximal, and $U_{e 3}$
small. This is in contrast to the quark sector, where the three independent
mixings are small or very small \cite{wol}.

$~$

Both features, small neutrino masses and large lepton mixings, can be accounted
for by the seesaw mechanism \cite{ss}. It requires only a modest extension of
the minimal standard model, namely the addition of a heavy fermionic singlet,
the right-handed neutrino $\nu_R$. In fact, this state permits to create a Dirac
mass $M_D$ for the neutrino, which is related to the electroweak symmetry
breaking and thus expected to be of the same order of quark or charged lepton
masses. Moreover, it permits also a Majorana mass $M_R$ for the right-handed
neutrino, which is not related to electroweak breaking and thus can be very
large. Then the effective neutrino mass matrix is given by
\ini
M_{\nu} \simeq M_D M_R^{-1} M_D=v^2 Y_D M_R^{-1} Y_D,
\fin
where $v$ is the v.e.v. of the standard Higgs field (electroweak breaking
scale), and $Y_D$ is the Yukawa coupling matrix. This seesaw mechanism
is called type I and of course gives small $M_{\nu}$, since $M_D$ is
suppressed by $M_D M_R^{-1}$.

$~$

There is also a triplet seesaw mechanism, which requires a heavy scalar
triplet $T$. In this case
\ini
M_{\nu}= M_L=Y_L v_L,
\fin
where $v_L$ is an induced v.e.v.,
\ini
v_L=\gamma \frac{v^2}{m_T}.
\fin
The sum of the two terms gives the type II seesaw formula \cite{ss2}
\ini
M_{\nu} \simeq M_D M_R^{-1} M_D +M_L.
\fin
Note that in the type I term there is a double matrix product, which can generate
large mixings from small mixings in $Y_D$, and also in $M_R$ \cite{smir}.
Instead in the type II (triplet) term, there is only one matrix, so that
large mixing is present or not, by hand...

Then we consider the type I term as fundamental and the triplet term as a kind
of pertubation.

$~$

Let us describe the effect on the mixing in the type II seesaw mechanism
\cite{df1}, using a model for mass matrices \cite{df2}, based on broken $U(2)$
horizontal symmetry and simple quark-lepton symmetry $M_e \sim M_d$,
$M_D \sim M_u$,
\ini
M_D \sim
\left( \begin{array}{ccc}
\la^{12} & \la^6 & \la^{10} \\
\la^6 & \la^4 & \la^4 \\
\la^{10} & \la^4 & 1
\end{array} \right)~m_t,
\fin
\ini
M_e \sim
\left( \begin{array}{ccc}
\la^{6} & \la^3 & \la^{5} \\
\la^3 & \la^2 & \la^2 \\
\la^{5} & \la^2 & 1
\end{array} \right)~m_b,
\fin
\ini
M_R \sim
\left( \begin{array}{ccc}
\la^{12} & \la^{10} & \la^6 \\
\la^{10} & \la^8 & \la^4 \\
\la^6 & \la^4 & 1
\end{array} \right)~m_R.
\fin
One could also adopt different mass matrices, keeping the analysis and possibly
the results quite similar.
The type I seesaw mechanism gives
\ini
M_{\nu}^I \sim
\left( \begin{array}{ccc}
\la^4 & \la^2 & \la^2 \\
\la^2 & 1 & 1 \\
\la^2 & 1 & 1
\end{array} \right)~\frac{m_t^2}{m_R},
\fin
corresponding to a normal hierarchy. Moreover we assume that the triplet term is
\ini
M_{\nu}^{II}=M_L=\frac{m_L}{m_R}M_R.
\fin
This form can be motivated within left-right and $SO(10)$ models. However,
in the mood of Ref.\cite{df2} we can think it to be generated by coupling to
the same flavon fields as $M_R$.

$~$

Now, since we do not write coefficients in mass matrices, consider a different
form of matrix (17),
\ini
M_{\nu}^I \simeq
\left( \begin{array}{ccc}
\la^4 & \la^2 & \la^2 \\
\la^2 & 1+\frac{\la^n}{2} & 1-\frac{\la^n}{2} \\
\la^2 & 1-\frac{\la^n}{2}  & 1+\frac{\la^n}{2}
\end{array} \right)~\frac{m_t^2}{m_R},
\fin
with $n=1,2,3,4$, which gives maximal $U_{\mu 3}$ but different $U_{e 2}$, according
to the value of $n$. In fact, for $n=4$ one has  $\sin \theta_{12}=1/\sqrt2$ and
$\epsilon=0$, that is the bimaximal mixing. For $n=3,2,1$ we get
$\sin \theta_{12} \simeq 0.68; 0.58; 0.25;$ respectively.

The contribution from $M_{\nu}^{II}$ will we parametrized by the ratio
\ini
k=\frac{m_{\nu}^{II}}{m_{\nu}^{I}} =\frac{m_L m_R}{v^2}=\gamma ~\frac{m_R}{m_T}
\fin
and leads to decrease of the mixing 2-3 and especially 1-2.

We consider numerical results in the following, but first the contribution
from $M_e$ should be taken into account. It is similar to the CKM matrix
in the Wolfenstein form and gives
\ini
\sin \theta_{e2} \simeq
\sin \theta_{12} -\frac{\la}{2},
\fin
\ini
\sin \theta_{\mu 3} \simeq \frac{1}{\sqrt2} \left( 1-\la^2 \right),
\fin
\ini
\sin \theta_{e3} \simeq -\frac{\la}{\sqrt2}.
\fin
However, we should also study the Georgi-Jarlskog (GJ) option \cite{gj}
\ini
M_e \sim
\left( \begin{array}{ccc}
\la^{6} & \la^3 & \la^{5} \\
\la^3 & -3 \la^2 & \la^2 \\
\la^{5} & \la^2 & 1
\end{array} \right)~m_b.
\fin
In this case we get
\ini
\sin \theta_{e2} \simeq
\sin \theta_{12}^{\nu} +\frac{\la}{6},
\fin
$$
\sin \theta_{23} \simeq \frac{1}{\sqrt2} \left( 1-\la^2 \right),
$$
\ini
\sin \theta_{13} \simeq \frac{\la}{3 \sqrt2}.
\fin
Finally we sum $M_{\nu}^{II}$ to $M_{\nu}^{I}$ and combine with the mixing
coming from $M_e$, and look for agreement with experimental values of mixings.

We find the following results:

Case $n=4$ requires $0 \le k<0.05$, or $0.08<k<0.18$ for the GJ option.

Case $n=3$ requires $0 \le k<0.04$, or $0.06<k<0.16$ for the GJ option.

Case $n=2$ is reliable only for the GJ choice with $0 \le k<0.10$.

Case $n=1$ is not reliable at all.

Note that the presence of zero means that the triplet term can be absent.
It is instead necessary for $n=4,3$ in the GJ option.

In particular, the difference between the observed quark-lepton complementarity
and the theoretical prediction based on realistic (GJ) quark-lepton symmetry
in cases $n=4,3$ could be ascribed to the triplet contribution within the type II
seesaw mechanism \cite{df3}. We predict
\ini
\theta_{e3} \simeq \lambda / 3 \sqrt2 \simeq 0.05,
\fin
which can be checked in future experiments.

\newpage

Now we consider a cosmological consequence of the seesaw mechanism,
namely the baryogenesis via leptogenesis \cite{fy,oth}. It aims to reproduce
the baryon asymmetry
\ini
\eta_B=\frac{n_B}{n_{\gamma}}= (6.1 \pm 0.2) \cdot 10^{-10}
\fin
by means of the out-of-equilibrium decays of the right-handed neutrinos
which generate a lepton asymmetry partially converted to baryon asymmetry
by electroweak sphaleron processes \cite{krs}.

The main formulas are
\ini
\eta_B \simeq 10^{-2} k_i \epsilon_i
\fin
where
\ini
\epsilon_1=\frac{3}{16 \pi} \frac{(Y_D^{\dg}Y_D)^2_{12}}{(Y_D^{\dg}Y_D)_{11}}
\frac{M_1}{M_2}
\fin
is a CP asymmetry related to the decay of the lightest right-handed neutrino, and
\ini
\epsilon_2=\frac{3}{16 \pi} \frac{(Y_D^{\dg}Y_D)^2_{23}}{(Y_D^{\dg}Y_D)_{22}}
\frac{M_2}{M_3}
\fin
to the decay of the next-to-lightest right-handed neutrino. These formulas in the
flavor basis.
For the matrix model considered above we get
\ini
\epsilon_1=\frac{3}{16 \pi} \lambda^{12} =2.4 \cdot 10^{-10}
\fin
\ini
\epsilon_2=\frac{3}{16 \pi} \lambda^{8} =1.5 \cdot 10^{-7}.
\fin
The dilution $k_i$ depends mainly on the parameter
$\tilde{m}_i =(M_D^{\dg} M_D)_{ii}/M_i$. In the present case we find
$\tilde{m}_2 \simeq m_1 \sim 10^{-4}$ eV, in the weak wash out regime,
$k_2 \sim 0.1$, so that
\ini
\eta_B \simeq 10^{-2} \cdot 10^{-1} \cdot 10^{-7}=10^{-10}
\fin
in agreement with the experimental value (28).
This result should be considered as an addendum to Ref.\cite{df4}, where
$\epsilon_2$ was not calculated.

$~$

In conclusion, the seesaw mechanism is able to explain the small neutrino
mass, the large and maximal mixing, and the cosmological baryon asymmetry.

$~$

$~$

We thank F. Buccella for discussions and support.


\begin{thebibliography}{100}

\newpage

\bibitem{sk} Y. Fukuda et al., Phys. Rev. Lett. 81 (1998) 1562

\bibitem{sno} Q.R. Ahmad et al., Phys. Rev. Lett. 89 (2002) 011301; 011302 

\bibitem{mns} Z. Maki, M. Nakagawa and S. Sakata, Prog. Theor. Phys. 28 (1962) 870

\bibitem{alt} G. Altarelli, Phys. Rep. 320 (1999) 295

\bibitem{wol} L. Wolfenstein, Phys. Rev. Lett. 51 (1983) 1945

\bibitem{ss} P. Minkowski, Phys. Lett. B 67 (1977) 421

R.N. Mohapatra and G. Senjanovic, Phys. Rev. Lett. 44 (1980) 912
 
\bibitem{ss2} G. Lazarides, Q. Shafi and C. Wetterich, Nucl. Phys. B 181 (1981) 287

R.N. Mohapatra and G. Senjanovic, Phys. Rev. D 23 (1981) 165

\bibitem{smir} A.Yu. Smirnov, Phys. Rev. D 48 (1993) 3264 

\bibitem{df1} D. Falcone, Int. J. Mod. Phys. A 21 (2006) 3015 

\bibitem{df2} D. Falcone, Phys. Rev. D 64 (2001) 117302

\bibitem{gj} H. Georgi and C. Jarlskog, Phys. Lett. B 86 (1979) 297

\bibitem{df3} D. Falcone, Mod. Phys. Lett. A 21 (2006) 1815

\bibitem{fy} M. Fukugita and T. Yanagida, Phys. Lett. B 174 (1986) 45

\bibitem{oth} M. Luty, Phys. Rev. D 45 (1992) 455

L. Covi, E. Roulet and F. Vissani, Phys. Lett. B 384 (1996) 169

W. Buchmuller and T. Yanagida, Phys. Lett. B 445 (1999) 399

D. Falcone and F. Tramontano, Phys. Rev. D 63 (2001) 073007

O. Vives, Phys. Rev. D 73 (2006) 073006

\bibitem{krs} V.A. Kuzmin, V.A. Rubakov and M.E. Shaposhnikov,
Phys. Lett. B 155 (1985) 36

\bibitem{df4} D. Falcone, Phys. Rev. D 68 (2003) 033002

\end{thebibliography}
\end{document}